\documentclass[aps,prd,superscriptaddress,showpacs, preprint]{revtex4-1}
\usepackage{graphicx}
\usepackage{epstopdf}
\usepackage{amsmath}
\usepackage{amsfonts}
\usepackage{amssymb}
\usepackage{latexsym}
\usepackage{hyperref}
\usepackage{xcolor}
\usepackage[colorinlistoftodos]{todonotes}

\usepackage{bm} 
\begin{document}

\def \L{\mathcal{L}}
\def \M{\mathcal{M}}
\def \O{\mathcal{O}}

\title{Fermion inter-particle potentials in 5D and a dimensional restriction prescription to 4D}

\author{L.P.R. Ospedal}\email{leoopr@cbpf.br}
\affiliation{Centro Brasileiro de Pesquisas F\'{i}sicas (CBPF), Rua Dr Xavier Sigaud 150, Urca, Rio de Janeiro, Brazil, CEP 22290-180}

\author{J.A. Helay\"{e}l-Neto}\email{helayel@cbpf.br}
\affiliation{Centro Brasileiro de Pesquisas F\'{i}sicas (CBPF), Rua Dr Xavier Sigaud 150, Urca, Rio de Janeiro, Brazil, CEP 22290-180}


\begin{abstract}

 This work sets out to compute and discuss effects of spin, velocity and dimensionality on inter-particle potentials systematically derived from gauge field-theoretic models. We investigate the interaction of fermionic particles by the exchange of a vector field in a parity-preserving description in five-dimensional $(5D)$ space-time.  A particular dimensional reduction prescription is adopted $-$ reduction by dimensional restriction $-$ and special effects, like a pseudo-spin dependence, show up in four dimensions $(4D)$. What we refer to as pseudo-spin shall be duly explained. The main idea we try to convey is that the calculation of the potentials in $5D$ and the consequent reduction to $4D$
exhibits new effects that are not present if the potential is calculated in $4D$ after the action has been reduced.
\end{abstract}

\pacs{   11.10.Kk, 11.15.Kc, 11.30.Er.}
\maketitle


\section{Introduction}
\indent

Field-theoretic models \`a la Kaluza-Klein have had a remarkable revival after the line of papers quoted in refs. \cite{1975_Scherk_Schwarz}-\cite{1981_Orzalesi_Pauri}, from which the activity known as Kaluza-Klein Supergravities  was boosted. An important question in connection with higher-dimensional models consists in computing quantum-mechanical effects. Two routes may be followed in connection with radiative corrections: in path $(i)$, one may carry out dimensional reduction by adopting some specific scheme and, then, once the reduction is carried out to some lower-dimensional space-time, quantum corrections are computed; route $(ii)$ proceeds in the reversed order: one computes the quantum effects directly in the higher-dimensional set-up of the model and compares, afterwards, with the quantum corrections computed in the lower-dimensional version of the model with the towers of massive fields  included. Procedures $(i)$ and $(ii)$ may not coincide. Actually, \'Alvarez and Faedo \cite{2006_Alvarez_Faedo} carefully discussed this issue and they found conditions so that  the two routes yield quantum-mechanically equivalent results.  

Back to 1983 and 1984, we point out a series of papers by Appelquist and Chodos \cite{1983_Alppequist_Chodos} \cite{1984_Chodos}, in which the authors consider the five-dimensional Kaluza-Klein model and compute the 1-loop effective potential for the extra component of the metric in 5D, attaining, therefore, the gravitational analogue of the Casimir Effect. Appelquist, Chodos and Myers \cite{1983_Alppequist_Chodos_Myers} also inspected how quantum effects may induce instabilities in the dimensional reduction process. Ever since, the issue of quantum corrections in higher dimensions and their residual effects in lower dimensions has become a very relevant activity in connection with models based on extra dimensions.


The main motivation of the present contribution lies on the problem of comparing results that follow if we adopt either of the routes, $(i)$ or $(ii)$, namely, quantum effects computed prior or after the dimensional reduction. We endeavor to tackle this question by considering a semi-classical aspect attainable from quantum field-theoretic models: inter-particle interaction potentials derived from the mediation of some intermediating particle. Our paper sets out to work out a spin- and velocity-dependent inter-particle potential between massive charged spin-$1/2$ particles in a five-dimensional formulation of parity-preserving Electrodynamics. (The usual Dirac mass term explicitly breaks parity symmetry in five dimensions. We here keep parity as a good symmetry and double the fermion representation, as it shall be clarified later on).

Even if no loop correction is computed, the tree-level one-scalar or one-photon exchange involves a quantum-mechanical object $-$ the causal propagator $-$ so that we get a semi-classical potential in five dimensions to be suitably reduced to four dimensions. This shall eventually trigger some new effect in four dimensions, inherent to the fact that quantization has already been introduced in five dimensions. The idea of a pseudo-spin, that will show up in four dimensions as a result of imposing parity conservation in five dimensions, is a consequence of considering the fundamental interaction taking place in five dimensions and the way to connect physics in four and five dimensions will be based on a procedure that we refer to as dimensional reduction by dimensional restriction. This shall be duly presented and discussed in Section \ref{S_restriction}. Had we first reduced the five-dimensional model, and then calculated the inter-particle potential, pseudo-spin interactions would not appear. 

The main point of our investigation is indeed to claim that an inter-particle potential in four space-time dimensions may exhibit extra spin effects that appear whenever we adopt the viewpoint that the quantum effects should be accounted for in five dimensions (where we consider that the fundamental physics takes place), rather than introducing quantum effects only after the dimensional reduction has been performed. In this scenario, deviations between theoretical results and experimental measurements could, in some cases, be originated from quantum-mechanical effects of physics that is processed in extra dimensions.

Even though truly fundamental physics in five dimensions should be associated to the five-dimensional anti-de Sitter  (in connection with the gauge/gravity correspondence) or de Sitter spaces (in connection with the accelerated expansion of the Universe), we understand that we are dealing with physical effects that are far from being sensitive to possible effects of the cosmological constant. We are bound to the scales of the Standard Model. Actually, we are considering electromagnetic effects, and the length scales involved in the physics we investigate are very far above the curvature of $\textrm{AdS}_5$ or $\textrm{dS}_5$. This our justification to consider that  the fundamental physics underneath our present investigation is consistent with $(1+4)$ Minkowski space-time.
 

Our paper is organized according to the following outline: in Section \ref{S_Methodology}, we review the Methodology for computing the spin- and velocity-dependent inter-particle potentials. In Section \ref{S_EM_5D}, we discuss the parity symmetry in $5D$ space-time for a massive Dirac spinor field and work out the potential for the  Maxwell Electrodynamics. In Section \ref{S_restriction}, we  propose a prescription for restrict the interaction from $5D$ to $4D$. Next, in Section \ref{S_Proca}, we also obtain the potential for the Proca Electrodynamics and show its asymptotic limits and restriction to $4D$. Finally, in Section \ref{S_Concluding}, we display our Concluding Comments. We shall adopt the  natural units $\hbar = c =1$.


\section{Methodology and Useful Results} 
\label{S_Methodology}
\indent 

We  consider an elastic scattering at tree-level of two particle with initial and final states given by $(E_{1,i}, \textbf{p}_{1,i} \, ; \, E_{2,i}, \textbf{p}_{2,i})$ and $(E_{1,f}, \textbf{p}_{1,f} \, ; \, E_{2,f}, \textbf{p}_{2,f})$, respectively. It is convenient to work in the center-of-mass (CM) reference frame with parametrization in terms of the two independent momentum: the transfer momentum, $\textbf{q} = \textbf{p}_{1,f} - \textbf{p}_{1,i} =  - ( \textbf{p}_{2,f} - \textbf{p}_{2,i} ) $, and the average momentum $ \textbf{p} = ( \textbf{p}_{1,i} + \textbf{p}_{1,f})/2 = (\textbf{p}_{2,i} + \textbf{p}_{2,f})/2 \,$. In this case, we have $ q^0 = 0$ and $ \textbf{q} \cdot \textbf{p} = 0$ to simplify the amplitude.

In the first Born approximation \cite{Maggiore}, the inter-particle potential in $4D$ space-time is obtained through Fourier Integral of the non-relativistic amplitude, 
\begin{eqnarray}
V = - \int \frac{d^3\textbf{q}}{(2\pi)^3} \,   e^{i \textbf{q}\cdot \textbf{r}} \, \M_{NR} \, \, ,
\label{def_pot_4D} \end{eqnarray}
where $ \M_{NR} $ is related to the  Feynman amplitude, $ \M$, by means of
\begin{eqnarray}\label{Amp_NR}
\M_{NR} = \frac{1}{\sqrt{2 E_{1,i}}} \frac{1}{\sqrt{2 E_{1,f}}} \frac{1}{\sqrt{2 E_{2,i}}} \frac{1}{\sqrt{2 E_{2,f}}} \, \M \, .
\label{def_Amp_NR} \end{eqnarray}
 We assume the metric $\eta_{\mu \nu} = \textrm{diag}(+,-,-,-) $.

In order to render this methodology more instructive and useful to the next Sections, let us  consider a particular case. For our purposes, it is convenient to work out the well-known electromagnetic interaction of fermions in $4D$, described by the Lagrangian density
\begin{equation}
\L = -\frac{1}{4} F_{\mu\nu}^2
+ \overline{\psi}  \left( i \gamma^\mu \partial_\mu - e \gamma^\mu A_\mu - m\right) \psi \, . \label{L_Maxwell}
\end{equation}


First of all, we need to exhibit the positive-energy solutions of the free Dirac equation, 
\begin{equation}
\left[ \gamma^\mu p_\mu - m \right] \psi(p) = 0 \, .
\label{Dirac_eq} \end{equation}
Using the decomposition in terms of two-component spinors, $\psi = (\xi , \chi)^t$, and taking the gamma-matrices in the Dirac representation, it is possible to eliminate $\chi$ and show that
\begin{equation}
\psi(p) = \frac{1}{\sqrt{E +m}} \, \left( \begin{array}{c} (E +m) \, \xi \\  {\bm \sigma} \cdot \textbf{p}  \, \; \xi 
\end{array} \right) \, , \label{psi_p_4D} \, \end{equation}
where we have normalized the spinor  such that $ \overline{\psi}(p) \psi(p) = 2m \, \xi^\dagger \xi$.

The basic spinor, $\xi$, may assume the following two values 
\begin{equation}
\xi =  \left( \begin{array}{c}  1 \\ 0  
\end{array} \right) \, \; , \, \; 
\left( \begin{array}{c}  0 \\ 1  
\end{array} \right) \, ,
\label{spin_up_down} \end{equation}
which refers to spin up and down configurations, respectively.

Here, we would like to fix some notations. Let us consider the possibility of a spin flip.
Thus, we shall use $\xi_i$ and $\xi_f$ to indicate the initial and final spin state of the fermion. For this reason, it is also convenient to define the contractions
\begin{equation}
\delta = \xi_f^\dagger \, \xi_i  \, \; , \; \,  
\langle \textbf{S} \rangle = \xi_f^\dagger \, \frac{ {\bm \sigma}}{2} \, \xi_i \, \; . 
\label{delta_spin_4D} \end{equation}
The previous expression is interpreted as the expectation value of the spin operator.

Now, we apply the Feynman rules for this scattering in the adopted CM frame, 
\begin{eqnarray} 
i \mathcal{M} & = & \overline{\psi}_{1}(\textbf{p} + \textbf{q}/2) \left\{ i e_1 
 \gamma^\mu  \right\} \psi_{1}(\textbf{p} - \textbf{q}/2) \langle A_\mu A_\nu\rangle \times \nonumber \\ \nonumber
& \times & \overline{\psi}_{2}(- \textbf{p} - \textbf{q}/2) \left\{ i e_2 \gamma^\nu  \right\} 
\psi_{2}(- \textbf{p} + \textbf{q}/2) = \nonumber \\
&=& \, - e_1 e_2 \, J_{(1)}^\mu \, \langle  A_\mu A_\nu \rangle \, 
J_{(2)}^\nu  
\, , \label{Rel_Amp_4D} \end{eqnarray}
where we are using the current $J^\mu = \overline{\psi} \gamma^\mu \psi $, and $\langle A_\mu A_\nu \rangle $ is the propagator in momentum space, 
\begin{equation}
\langle A_\mu A_\nu \rangle = - \frac{i}{q^2} \left[ \eta_{\mu\nu} + (\alpha - 1) \frac{q_\mu q_\nu}{q^2} \right]  \, , \label{prop_Maxwell}
\end{equation}
which is obtained after including the gauge-fixing term, $\frac{-1}{2 \alpha}(\partial_\mu A^\mu)^2 $ , to the Lagrangian in eq. \eqref{L_Maxwell}.

With the current conservation, $q^\mu J_\mu = 0$, and $q^0 =0$ in eq. \eqref{Rel_Amp_4D},  the non-relativistic amplitude, eq. \eqref{def_Amp_NR}, can be written as 
\begin{equation}
\M_{NR} = - \frac{e_1 e_2}{ \textbf{q}^{2}  } \, \frac{J_{(1)}^{\mu} J_{(2) \,\mu}}{ (2E_1) (2E_2) } \, . \label{M_NR_4D} \end{equation}

One important step of this computation is to declare which approximation we are dealing with. Throughout this work, we consider corrections up to $\O (|\textbf{p}^2|/m^2)$ in the amplitude, without counting the factor $1/\textbf{q}^2$ in the previous equation.

Now, we  study the currents in order to  obtain an approximation to this amplitude. For the particle$-1$, we take the spinor solution, eq. \eqref{psi_p_4D}, and consider $E_{1,f} = E_{1,i} = E_1 \approx m_1 + \frac{1}{2 m_1} \, \left( \textbf{p}^2 + \frac{\textbf{q}^2}{4} \right) $ such that
\begin{eqnarray}
J_{(1)}^0 & \approx & 2 m_1  \, \delta_1 + \frac{1}{ m_1} \, \biggl[ \,   \textbf{p}^{2} \, \delta_1  
+ i \, \left( \textbf{q} \times \textbf{p} \right) \cdot \langle \textbf{S}_1 \rangle  \biggr] \, ,\label{J0_4D} \end{eqnarray}
\begin{equation}
J_{(1)}^i  \approx  2   \textbf{p}_i \, \delta_1  - 2i  \, \epsilon_{ijk} \, \textbf{q}_j \, \langle \textbf{S}_{1,k} \rangle \, .\label{Ji_4D} \end{equation}  

The current $J^\mu_{(2)}$ is obtained by taking the following prescription in the  $J^\mu_{(1)}$: $ \textbf{ q} \rightarrow - \textbf{ q} $, $ \textbf{ p} \rightarrow - \textbf{ p} $ and changing the label $1 \rightarrow 2$.

From these considerations, one could check that
\begin{eqnarray}
\frac{J_{(1)}^{\mu} J_{(2) \,\mu}}{ (2E_1) (2E_2) } & \approx &  \delta_1 \delta_2 \left[ \left( 1 + \frac{\textbf{p}^{2}}{m_1 m_2}  \right) \right. 
- \left. \frac{1}{8} \left( \frac{1}{m^2_1} + \frac{1}{m^2_2} \right) \,  \textbf{q}^{2}   \right]
+ \nonumber \\
& + & i \textbf{q} \cdot \left\{ \textbf{p} \times \left[ 
\delta_1 \langle \textbf{S}_2 \rangle \left(\frac{1}{2 m_2^2} + \frac{1}{m_1 m_2} \right) + 
1 \leftrightarrow 2 \right] \right\} + \nonumber\\
& - &   \frac{1}{m_1 m_2} \, \left[ \, \textbf{q}^{2} \,
 \langle \textbf{S}_1 \rangle \cdot \langle \textbf{S}_2 \rangle - \left( \textbf{q} \cdot \langle \textbf{S}_1 \rangle \right)
\left( \textbf{q} \cdot \langle \textbf{S}_2 \rangle \right)
\, \right] \, . \label{JJ_4D} \end{eqnarray}

Once established the non-relativistic amplitude, eq. \eqref{M_NR_4D} with eq. \eqref{JJ_4D}, we use the prescription described in eq. \eqref{def_pot_4D}, i.e., we carry out the Fourier integral. For this calculation, we only need the massless limit of eqs. \eqref{int_4D}-\eqref{int_4D_qq} of the Appendix \ref{Apendice}. Then, the inter-particle potential is given by
\begin{eqnarray} 
V^{\textrm{Maxwell}} & = & e_1 e_2 \left\{  \frac{\delta_1\delta_2}{4 \pi r}  \left[ 1 +  \, \frac{\textbf{p}^{2}}{m_1 m_2} \right]   - \frac{\delta_1\delta_2}{8} \,  \left( \frac{1}{m_1^2} + \frac{1}{m_2^2} \right) \, \delta^{3}(\textbf{r}) + \right. \nonumber\\ 
&-&  \frac{2}{3} \frac{\langle \textbf{S}_1 \rangle \cdot \langle \textbf{S}_2 \rangle}{m_1 m_2} \, \delta^{3}(\textbf{r}) + \frac{\textbf{Q}_{ij}}{4 \pi r^3} \,\frac{ \,
 \langle \textbf{S}_{1,i} \rangle \, \langle \textbf{S}_{2,j} \rangle  }{m_1 m_2} +  \nonumber\\ 
&-&  \left. \frac{\textbf{L}}{4 \pi r^3} \,   \cdot \left[  
\delta_1 \langle \textbf{S}_2 \rangle \left(\frac{1}{2 m_2^2} + \frac{1}{m_1 m_2} \right) + 1 \leftrightarrow 2 \right]  \right\} \, ,
\label{V_4D_Maxwell} \end{eqnarray}
where we defined the angular momentum, $ \textbf{L} = \textbf{r} \times \textbf{p} $, and  
quadrupole (or dipole-dipole) tensor $ \textbf{Q}_{ij} = \delta_{ij} - 3 \, \frac{ \textbf{x}_i \textbf{x}_j }{r^2}  $.

The first contribution is the usual Coulomb interaction $(\sim 1/4 \pi r)$, which is the dominant term at large distances. Next, we have a velocity-dependent term, here parametrized in terms of the average momentum $ \textbf{p}$. It also has a spin-orbit coupling $\textbf{L} \cdot \textbf{S}$, quadrupole interaction and contact terms, i.e.,  ones with Dirac delta $\delta^{3}(\textbf{r})$. Due our approximations, we do not have higher multipole contribution than quadrupole. This result coincides with the one obtained in refs. \cite{Pedro_kim} \cite{Gustavo_Pedro}.  We shall see in the following Sections that the calculation of the inter-particle potential in $5D$ space-time follows similar procedures as the ones presented in this particular case.


\section{Maxwell Electrodynamics in 5D}
\label{S_EM_5D}
\indent

 The properties of the inter-particle interaction potentials in arbitrary dimensions has been already established in the literature  for many situations, see refs. \cite{Accioly_ED_EM}-\cite{Accioly_renor_origin}. However, there is a lack of the attention to the study related to spin contributions for space-times with extra dimensions. In this Section, we pursue an investigation of the spin as well as velocity-dependent interactions in space-time with one extra dimension. Initially, we concentrate our efforts in the Maxwell Electrodynamics in $5D$  space-time. Keeping in mind that Electromagnetism is also parity-invariant in $5D$, we start by studying how to implement the parity transformation on massive Dirac fermions.

Let us initiate by fixing other conventions. In $5D$ Minkowski space-time, we adopt the metric $\eta_{\hat{\mu} \hat{\nu}} = \textrm{diag}(+,-,-,-,-) $, where $ \hat{\mu}, \hat{\nu} = (0,i,4)$ with $i=(1,2,3)$. One possible choice to satisfy the Clifford algebra, 
$ \left\{ \gamma^{\hat{\mu}} , \gamma^{\hat{\nu}} \right\} = 2 \eta^{\hat{\mu} \hat{\nu}} $, is to take
$ \gamma^{\hat{\mu}} = \left( \gamma^{\mu} , \gamma^4 \equiv i \gamma_5 \right) $, where $ \gamma_5 = i \gamma^0 \gamma^1 \gamma^2 \gamma^3 $ and $\gamma^\mu $ satisfies the Clifford algebra in $4D$. Another possibility is $ \gamma^{\hat{\mu}} = \left( i \gamma_5 \gamma^{\mu} , \gamma^4 \equiv i \gamma_5 \right) $. We consider the first one, which will be more convenient to the evaluations in the non-relativistic limit.

The Lagrangian for a massive Dirac spinor field in $5D$ is given by 
\begin{equation}
\L = \overline{\psi} \, i \gamma^{\hat{\mu}} \partial_{\hat{\mu}} \psi - m \, \overline{\psi} \psi
\, . \label{L_Dirac} \end{equation}
 
We define the parity transformation in $5D$ spacetime as $x_0' = x_0 \, , \, \textbf{x}' = - \textbf{x} \, $ and $ \, \textbf{x}_4' = \textbf{x}_4$. Thus, we maintain the usual transformation in $4D$ and the extra-dimension, $\textbf{x}_4$, stays unaltered in order to have a discrete transformation. Let us propose the following parity transformation for the spinor field
\begin{equation}
\psi'(x') = P \, \psi(x) \, .
\label{trans_psi} \end{equation}

Now, we would like to find an explicitly form for the  matrix $P$. We start by imposing the invariance of the massless term in eq. \eqref{L_Dirac}. Using  eq. \eqref{trans_psi} and $\overline{\psi'} = \psi'^\dagger \gamma^0 = \overline{\psi} \, \gamma^0 P^\dagger \gamma^0 $, one could obtain the relations
\begin{equation}
P^\dagger = P^{-1} \, \; , \, \; \gamma^0 \gamma^i \, P = - P \, \gamma^0 \gamma^i
\, \; , \, \; \gamma^0 \gamma^4 \,  P = P \, \gamma^0 \gamma^4 \, .
\end{equation}
which provide us 
\begin{equation}
P = i \, \gamma^1 \gamma^2 \gamma^3 \, .
\label{transf_P} \end{equation}

The factor $i$ is just for future convenience. However, we have not finished yet, we need to consider the transformation of the mass term in eq. \eqref{L_Dirac}. From the above result, we find that
\begin{equation}
m \, \overline{\psi'} \, \psi' = m \, \overline{\psi} \, \gamma^0 P^\dagger \gamma^0 P \, \psi = - m \, \overline{\psi} \, \psi
\label{trans_mass} \end{equation}
so, the mass term breaks the parity symmetry in $5D$.

One way to circumvent this problem is to double the spinor field representation. A similar proposal was taken in ref. \cite{QED_3}, in the context of $3D$ Minkowski space-time, also to conciliate the parity symmetry with massive fermions. Another possibility is to modify the mass term, as done in ref. \cite{kocinski}, but we will not follow this path here. Therefore, we define a doubled spinor field:
 \begin{equation}
\Psi =  \left( \begin{array}{c}  \psi \\ \chi  
\end{array} \right) \, .\label{def_Psi} \end{equation}

We also represent the gamma-matrices as
\begin{equation}
\Gamma^{\hat{\mu}} =  \left( \begin{array}{cc} \gamma^{\hat{\mu}}  & 0 \\ 0 & - \gamma^{\hat{\mu}}
\end{array} \right) \label{def_Gamma} \, ,\end{equation}
then the Dirac conjugate of $\Psi$ takes the form $ \overline{\Psi} = \Psi^\dagger \Gamma^0 = \left( \overline{\psi} \, , \, - \overline{\chi} \right) $, with $\overline{\psi} = \psi^\dagger \gamma^0$ and  $\overline{\chi} = \chi^\dagger \gamma^0$.

The Dirac Lagrangian for the doubled spinor field is given by
\begin{eqnarray} \L &=& \overline{\Psi} \, i \, \Gamma^{\hat{\mu}} \, \partial_{\hat{\mu}} \Psi - m \overline{\Psi} \, \Psi 
= \nonumber\\
&=& \overline{\psi} \, i \gamma^{\hat{\mu}} \, \partial_{\hat{\mu}} \psi + \overline{\chi} \, i \gamma^{\hat{\mu}} \, \partial_{\hat{\mu}} \chi 
- m \, \overline{\psi} \, \psi + m \, \overline{\chi} \, \chi
\, .\label{L_2x} \end{eqnarray}

After  these considerations, if we implement the parity transformation on $\Psi$ as
\begin{equation}
\Psi' =  \left( \begin{array}{cc} 0 & P \\ P & 0 \end{array} \right) 
\left( \begin{array}{c} 
\psi \\ \chi \end{array} \right)
\label{transf_P_2x} \, ,\end{equation}
then, it is possible to show that the Lagrangian of eq. \eqref{L_2x} is parity-invariant, since the transformation exchanges
\begin{equation}  \overline{\psi} \, i \gamma^{\hat{\mu}} \, \partial_{\hat{\mu}} \psi \longleftrightarrow \overline{\chi} \, i \gamma^{\hat{\mu}} \, \partial_{\hat{\mu}} \chi \, \; \, , \, \; \, 
 - m \, \overline{\psi} \, \psi \longleftrightarrow m \, \overline{\chi} \, \chi \, .\end{equation}

 It is worthy to mention that, similar to $3D-$case \cite{QED_3} (with $\tau_3-$QED), we could introduce other symmetries in the doubled field formalism. These possibilities shall be discussed in more details in the Concluding Comments.

Now, we are ready to start the steps for computing the inter-particle potential in $5D$. We shall follow the prescription described in the Section \ref{S_Methodology}. First, we need to obtain the free positive-energy solution of the Dirac equation, 
\begin{equation} \left( \Gamma^{\hat{\mu}} p_{\hat{\mu}} - m \right) \Psi(p) = 0 \, ,\end{equation}
which is equivalent to
\begin{equation} \left( \gamma^{\hat{\mu}} p_{\hat{\mu}} - m \right) \psi(p) = 0 \, \; \, , \, \; \, \left( \gamma^{\hat{\mu}} p_{\hat{\mu}} + m \right) \chi(p) = 0 \, .
\label{eq_psi_chi} \end{equation}

Then, we consider the decomposition
\begin{equation}
\psi =  \left( \begin{array}{c}  \xi \\ \varphi  
\end{array} \right) \, \; , \, \; 
\chi =  \left( \begin{array}{c}  \lambda \\ \zeta
\end{array} \right) \label{psi_chi_2x} \, , \end{equation}
where $ \xi , \varphi , \lambda$ and $\zeta$ are two-component spinors. Using eqs. \eqref{eq_psi_chi},  one can  eliminate $ \varphi , \lambda $ and  the spinors reduce to
\begin{equation}
\psi(p) = \frac{1}{\sqrt{E+m}} \, \left( \begin{array}{c} (E+m) \, \xi \,  \\ \left( {\bm \sigma} \cdot \textbf{p} - i \, \textbf{p}_4 \right) \, \xi 
\end{array} \right) \label{psi_p} \, , \end{equation}

\begin{equation}
\chi(p) = \frac{1}{\sqrt{E+m}} \, \left( \begin{array}{c}  
\left( {\bm \sigma} \cdot \textbf{p} + i \, \textbf{p}_4 \right)\, \zeta \\ (E+m) \, \zeta \end{array} \right) \label{chi_p} \, .\end{equation}

The two spinors above were normalized such that the doubled spinor field, eq. \eqref{def_Psi}, satisfies $ \overline{\Psi}(p) \Psi(p) = 2m \left(  \xi^\dagger \xi + \zeta^\dagger \zeta \right) $. Furthermore, they differ  by a minus sign in the extra dimension term, i.e., in the $\textbf{p}_4$ term. This sign is essentially to maintain the parity symmetry in $5D$ and  will play an important consequence in the spin interactions present in our $5D$ scenario. We expected to get more interactions in the doubled field formalism and the parity breaking case is recovered by taking $\zeta = 0$.

Since we are dealing with  $\xi$ and $\zeta$, we  introduce a label in the contractions given in eq. \eqref{delta_spin_4D}, so we define
\begin{equation}
\delta_\xi = \xi_f^\dagger \, \xi_i \, \; , \, \; \delta_\zeta = \zeta_f^\dagger \, \zeta_i \, \; , \, \; \langle \textbf{S} \rangle_{\xi} = \xi_f^\dagger \, \frac{ {\bm \sigma}}{2} \, \xi_i \, \; , \, \; \langle \textbf{S} \rangle_{\zeta} = \zeta_f^\dagger \, \frac{ {\bm \sigma}}{2} \, \zeta_i \, \; .
\label{exp_values} \end{equation}

Having established the spinors solutions, we turn to the calculation of the doubled field vector current,
\begin{equation}
J^{\hat{\mu}} = \overline{\Psi} \, \Gamma^{\hat{\mu}} \, \Psi = 
\overline{\psi} \, \gamma^{\hat{\mu}} \, \psi +
\overline{\chi} \, \gamma^{\hat{\mu}} \, \chi
\, .\label{double_V_current} \end{equation}

Inserting eqs. \eqref{psi_p} and \eqref{chi_p} in eq. \eqref{double_V_current} and using the adopted CM frame, 
one could show that the components of the current of the particle $1$ assume the form
\begin{eqnarray}
J_{(1)}^0 & = & 2 m_1 \left( \delta_{\xi,1} + \delta_{\zeta,1} \right) + \frac{1}{ m_1} \, \biggl\{ \,  \left( \delta_{\xi,1} + \delta_{\zeta,1} \right) \left( \textbf{p}^{2} + \textbf{p}^{2}_4 \right) + \nonumber\\ 
&+&  i \, \left( \textbf{q} \times \textbf{p} \right) \cdot \left[ \langle \textbf{S}_1 \rangle_{\xi} + \langle \textbf{S}_1 \rangle_\zeta \right]
  +   i \, \textbf{q}_4 \,  \textbf{p} \cdot \left[ \, \langle \textbf{S}_1 \rangle_{\xi} - \langle \textbf{S}_1 \rangle_\zeta \, \right]  + \nonumber\\  
&-& i \, \textbf{p}_4  \, \textbf{q} \cdot \left[ \, \langle \textbf{S}_1 \rangle_{\xi} - \langle \textbf{S}_1 \rangle_\zeta \, \right]   \, \biggr\} \, ,
\label{J0_2x} \end{eqnarray}
\begin{eqnarray}
J_{(1)}^i  &=&  2 \left( \delta_{\xi,1} + \delta_{\zeta,1} \right)  \textbf{p}_i - 2i \, \epsilon_{ijk} \, \textbf{q}_j \, \left[ \, \langle \textbf{S}_{1,k} \rangle_{\xi} + \langle \textbf{S}_{1,k} \rangle_\zeta \, \right] + \nonumber\\
&+& 2i \, \textbf{q}_4 \, \left[ \, \langle \textbf{S}_{1,i} \rangle_{\xi} - \langle \textbf{S}_{1,i} \rangle_\zeta \, \right] 
\, ,\label{Ji_2x} \\  \nonumber \\ 
J_{(1)}^4  &=&  2 \left( \delta_{\xi,1} + \delta_{\zeta,1} \right) \textbf{p}_4 - 2i \, \textbf{q} \cdot \left[ \, \langle \textbf{S}_{1} \rangle_{\xi} - \langle \textbf{S}_{1} \rangle_\zeta \, \right] \, .
\label{J4_2x} \end{eqnarray} 

Before going to the amplitude, it is interesting to look  carefully at these equations and their parity transformations. 
According to eq. \eqref{transf_P_2x} and the spinor solution, eqs. \eqref{psi_p} and \eqref{chi_p},
one can check that the parity transformation of the components are $ \xi' = \zeta$ and $\zeta' = - \xi $. That is the reason we put the factor $i$ in $P= i \gamma^1 \gamma^2 \gamma^3$, i.e., to get a real transformation. 
Since $\textbf{q}' = - \textbf{q} $, $\textbf{q}_4' = \textbf{q}_4$,
$\textbf{p}' = - \textbf{p}$ and $\textbf{p}_4' = \textbf{p}_4$,
we note some specific linear combinations of the spins in order to keep the  parity property of the vector in $5D$. For example, in the second term of eq. \eqref{J4_2x}, we have $ -2i \,\textbf{q}' \cdot \left[ \, \langle \textbf{S}_{1} \rangle_{\xi}' - \langle \textbf{S}_{1} \rangle_\zeta' \, \right] = -2i \, \textbf{q} \cdot \left[ \, \langle \textbf{S}_{1} \rangle_{\xi} - \langle \textbf{S}_{1} \rangle_\zeta \, \right]$, which is consistent with $J_{(1)}^{'4} = J_{(1)}^4$. A similar argument holds for the other terms. Therefore, it is suggestive to define 
\begin{equation}
\langle \textbf{S}^{\pm} \rangle =   \langle \textbf{S} \rangle_{\xi} \pm \langle \textbf{S} \rangle_\zeta
\label{Spin_2x} \, . \end{equation}
which can be understood as the bilinears below:
\begin{eqnarray}
\langle \textbf{S}^{\pm} \rangle = 
 \left( \begin{array}{cc} \xi_f^\dagger \, , \, \zeta_f^\dagger  \end{array} \right) \, \textbf{S}^{\pm}
 \left( \begin{array}{c} \xi_i \\ \zeta_i  \end{array} \right)  \label{media_spin_pm} \, \end{eqnarray}
 where
 \begin{equation}
\textbf{S}^{\pm} = \frac{1}{2} \, \left( \begin{array}{cc}  {\bm \sigma} & 0 \\ 0 & \pm { \bm \sigma} \end{array} \right) \label{spin_pm} \, . \end{equation}
 
The $ \langle \textbf{S}^+ \rangle $  can be interpreted as a expectation value of the spin, because the operator $\textbf{S}^+$ satisfies the $SU(2)$ algebra, $ \left[ \textbf{S}^+_i , \textbf{S}^+_j \right] = i \epsilon_{ijk} \textbf{S}^+_k$, and its expectation value is even under parity, $\langle \textbf{S}^{+} \rangle' = \langle \textbf{S}^{+} \rangle $, as true spin should be. On the other hand, the operator $ \textbf{S}^-$ and its expectation value do not satisfy these properties. Once $ \textbf{S}^-$ is formed by two spin ${ \bm \sigma}/2$ and under parity satisfies $\langle \textbf{S}^{-} \rangle' = - \langle \textbf{S}^{-} \rangle $, we shall call it  pseudo-spin. We  highlight that the pseudo-spin we introduce here is not the same as the pseudo-spin that appears in other contexts; for example, in Condensed Matter Systems \cite{pseudospin_CM} and Nuclear Physics \cite{pseudospin_NP_1} \cite{pseudospin_NP_2}.

It is also convenient to define $ \Delta = \delta_\xi + \delta_\zeta $, which is parity-invariant, $\Delta' = \Delta$. 

After  these definitions, we can recast the components of the vector current, eqs. \eqref{J0_2x}-\eqref{J4_2x}, as follows
\begin{eqnarray}
J_{(1)}^0 & = & 2 m_1 \Delta_1 + \frac{1}{ m_1} \, \biggl[ \, \Delta_1 \left( \textbf{p}^{2} + \textbf{p}^{2}_4 \right) 
+ i \, \left( \textbf{q} \times \textbf{p} \right) \cdot \langle \textbf{S}_1^{+} \rangle  + \nonumber\\ 
& + &  i \, \textbf{q}_4 \, \left( \textbf{p} \cdot \langle \textbf{S}_1^{-} \rangle  \right) 
- i \, \textbf{p}_4 \left( \textbf{q} \cdot \langle \textbf{S}_1^{-} \rangle  \right) \, \biggr] \, ,
\label{J0_2x_red} \end{eqnarray}
    
\begin{equation}
J_{(1)}^i  =  2 \Delta_1 \, \textbf{p}_i - 2i \, \epsilon_{ijk} \, \textbf{q}_j \, \langle \textbf{S}_{1,k}^{+} \rangle
+ 2i \, \textbf{q}_4 \, \langle \textbf{S}_{1,i}^{-} \rangle \, , \label{Ji_2x_red} \end{equation}     
    
\begin{equation}
J_{(1)}^4  =  2 \Delta_1 \, \textbf{p}_4 - 2i \, \textbf{q} \cdot \langle \textbf{S}_1^{-} \rangle
\, , \label{J4_2x_red} \end{equation}
  
They exhibit all the contributions of the components in $4D$, see eqs. \eqref{J0_4D} and \eqref{Ji_4D}, and new terms associated with extra-dimension.   

We can now proceed to evaluating  the amplitude in the context of Maxwell Electrodynamics in $5D$. In a similar way as done in the  Section \ref{S_Methodology}, one can show that  
\begin{equation}
\M_{NR}^{5D} = - \frac{g_1 g_2}{ \textbf{q}^{2} + \textbf{q}_4^{2} } \, \frac{J_{(1)}^{\hat{\mu}} J_{(2) \,\hat{\mu}}}{ (2E_1) (2E_2) } \, ,\label{M_NR_5D} \end{equation}
where $g_{1(2)}$ denotes the coupling constant in $5D$.

After some manipulations, we find that
%
\begin{eqnarray}
\frac{J_{(1)}^{\hat{\mu}} J_{(2) \,\hat{\mu}}}{ (2E_1) (2E_2) } & \approx &  \Delta_1 \Delta_2 \left[ \left( 1 + \frac{\textbf{p}^{2}}{m_1 m_2} + \frac{\textbf{p}_4^{2}}{m_1 m_2} \right) \right. + \nonumber \\
&-& \left. \frac{1}{8} \left( \frac{1}{m^2_1} + \frac{1}{m^2_2} \right) \, \left( \textbf{q}^{2} + \textbf{q}^{2}_4 \right) \right]
+ \nonumber \\
& + & i \textbf{q} \cdot \left\{ \textbf{p} \times \left[ 
\Delta_1 \langle \textbf{S}^{+}_2 \rangle \left(\frac{1}{2 m_2^2} + \frac{1}{m_1 m_2} \right) \right. 
\right] + \nonumber\\
&-& \left. \left. \textbf{p}_4 \left[ \Delta_1 \langle \textbf{S}^{-}_2 \rangle \left(\frac{1}{2 m_2^2} + \frac{1}{m_1 m_2} \right) \right] + 1 \leftrightarrow 2  \right\} \right. + \nonumber \\
&+& i \, \textbf{q}_4 \, \textbf{p} \, \left[ \Delta_1 \langle \textbf{S}^{-}_2 \rangle \left(\frac{1}{2 m_2^2} + \frac{1}{m_1 m_2} \right) + 1 \leftrightarrow 2  \right] + \nonumber \\
&+& \frac{\textbf{q}_4 \, \textbf{q}}{m_1 m_2} \, \cdot \left[ \, \left( \langle \textbf{S}^{+}_1 \rangle \times \langle \textbf{S}^{-}_2 \rangle \right) + \left( \langle \textbf{S}^{+}_2 \rangle \times \langle \textbf{S}^{-}_1 \rangle \right) \, \right]
+ \nonumber \\
& - &   \frac{1}{m_1 m_2} \, \left[ \, \textbf{q}^{2} \,
 \langle \textbf{S}^{+}_1 \rangle \cdot \langle \textbf{S}^{+}_2 \rangle  
 + \textbf{q}^{2}_4 \, \langle \textbf{S}^{-}_1 \rangle \cdot \langle \textbf{S}^{-}_2 \rangle \,
 \right] + \nonumber\\ 
&+& \frac{1}{m_1 m_2} \, \left[ \, \left( \textbf{q} \cdot \langle \textbf{S}^{+}_1 \rangle \right)
\left( \textbf{q} \cdot \langle \textbf{S}^{+}_2 \rangle \right) - \left( \textbf{q} \cdot \langle \textbf{S}^{-}_1 \rangle \right) \left( \textbf{q} \cdot \langle \textbf{S}^{-}_2 \rangle \right) \, \right] \, .
\label{JJ_5D} \end{eqnarray}

Finally, we only need to compute the Fourier integral,
\begin{eqnarray}\label{energia}
V_{5D} = - \int \frac{d^4 \textbf{q}}{(2\pi)^4} \,   e^{i \textbf{q}\cdot \textbf{R}} \, \M^{5D}_{NR} \, .
\label{def_pot_5D} \end{eqnarray}

As explained in the Appendix, we shall use $\textbf{R}$ to denote the $4D$ euclidean vector, so we  avoid confusion with $\textbf{r}$, used for the $3D-$case. Therefore, using the massless Fourier integrals in $4D$, given by eqs. \eqref{int_massless}-\eqref{int_qq_massless}, we obtain 
\begin{eqnarray} 
V^{\textrm{Maxwell}}_{5D} & = & g_1 g_2 \left\{ \frac{\Delta_1\Delta_2 }{4 \pi^2 R^2}  \,  \left( 1 + \frac{ \textbf{p}^{2}}{m_1 m_2} + \frac{ \textbf{p}_4^{2}}{m_1 m_2} \right) \right. + \nonumber\\
&-&  \frac{\Delta_1\Delta_2}{8}   \left( \frac{1}{m_1^2} + \frac{1}{m_2^2} \right)  \delta^{4}(\textbf{R}) + \nonumber\\
&-& \frac{1}{2 m_1 m_2} \left[ \, \langle \textbf{S}^{+}_1 \rangle \cdot \langle \textbf{S}^{+}_2 \rangle + 
\langle \textbf{S}^{-}_1 \rangle \cdot \langle \textbf{S}^{-}_2  \rangle \, \right] \delta^{4}(\textbf{R}) + \nonumber\\ 
&-&  \frac{\left( \textbf{r} \times \textbf{p} \right)}{2 \pi^2 R^4} \,   \cdot \left[  
\Delta_1 \langle \textbf{S}^{+}_2 \rangle \left(\frac{1}{2 m_2^2} + \frac{1}{m_1 m_2} \right) + 1 \leftrightarrow 2 \right]   + \nonumber\\
& - &  \frac{ \left( \textbf{x}_4 \, \textbf{p} - \textbf{p}_4 \, \textbf{r} \right)}{2 \pi^2 R^4} \, \cdot \left[ \Delta_1 \langle \textbf{S}^{-}_2 \rangle \left(\frac{1}{2 m_2^2} + \frac{1}{m_1 m_2} \right) + 1 \leftrightarrow 2 \right]   + \nonumber\\
& - &  \frac{2 }{ \pi^2 m_1 m_2}  \, \frac{ \textbf{x}_4 \, \textbf{r}}{R^6} \cdot \biggl[ \left( \langle \textbf{S}^{+}_1 \rangle \times \langle \textbf{S}^{-}_2 \rangle \right) + \left(  \langle \textbf{S}^{+}_2 \rangle \times \langle \textbf{S}^{-}_1 \rangle \right)  \biggr]  + \nonumber\\ 
& - &  \frac{1}{\pi^2 R^4} \frac{1}{m_1 m_2} \, \left[ \left( 1 - \frac{2 r^2}{R^2} \right) \biggl(   \langle \textbf{S}^{+}_1 \rangle \cdot \langle \textbf{S}^{+}_2 \rangle - 
\langle \textbf{S}^{-}_1 \rangle \cdot \langle \textbf{S}^{-}_2  \rangle  \biggr)  + \right. \nonumber\\
& + & \left. \left. \frac{2}{R^2}\, \left( \langle \textbf{S}^{+}_1  \rangle \cdot \textbf{r} \right) \left( \langle \textbf{S}^{+}_2  \rangle \cdot \textbf{r} \right) - \frac{2}{R^2} \left( \langle \textbf{S}^{-}_1  \rangle \cdot \textbf{r} \right) \left( \langle \textbf{S}^{-}_2  \rangle \cdot \textbf{r} \right) \,  \right] \right\} \, .
\label{EM_5D} \end{eqnarray}


 The dominant contribution at large distances is given by the first term $( \sim 1/4\pi^2 R^2)$. Similar to $3D-$case, see eq. \eqref{V_4D_Maxwell}, we obtain a spin-orbit coupling, ${\bm L} \cdot {\bm S^+}$, where ${\bm L} = {\bm r} \times {\bm p}$ is the $3D-$angular momentum, and an extra-component that couples with pseudo-spin, namely, the term related to $ \left( \textbf{x}_4 \, \textbf{p} - \textbf{p}_4 \, \textbf{r} \right) \cdot {\bm S}^-$.   As anticipated at the beginning, the doubled spinor  formalism provides  new interactions in $5D$ than the usual (non-doubled) formalism. For example, we observe a non-trivial coupling between spin and pseudo-spin of the type $ \langle \textbf{S}^{+} \rangle \times \langle \textbf{S}^{-} \rangle $ with a ${\bm x}_4 {\bm r}/R^6$ power-law decay.
We can check that this contribution only exists in the parity-invariant case. For instance, let us examine  its (pseudo)-spin dependence:
\begin{equation}  \left( \, \langle \textbf{S}^{+}_1 \rangle \times \langle \textbf{S}^{-}_2 \rangle \, \right) + \left( \, \langle \textbf{S}^{+}_2 \rangle \times \langle \textbf{S}^{-}_1 \rangle \, \right)  = 2 \left( \, \langle \textbf{S}_1 \rangle_\zeta \times \langle \textbf{S}_2 \rangle_\xi \, \right)  + 2 \left(  \, \langle \textbf{S}_2 \rangle_\zeta \times \langle \textbf{S}_1 \rangle_\xi \, \right) \, ,\end{equation} 
 so, in the parity breaking case, we  take  $\zeta =0$, which implies $ \langle \textbf{S} \rangle_\zeta = 0$ and leads to a trivial contribution. A similar argument holds for the last term of the potential, the quadrupole-like interaction, since
\begin{eqnarray}  \left( \langle \textbf{S}^{+}_1  \rangle \cdot \textbf{r} \right) \left( \langle \textbf{S}^{+}_2  \rangle \cdot \textbf{r} \right) - \left( \langle \textbf{S}^{-}_1  \rangle \cdot \textbf{r} \right) \left( \langle \textbf{S}^{-}_2  \rangle \cdot \textbf{r} \right)  
&=& 2 \left( \langle \textbf{S}_1  \rangle_\zeta \cdot \textbf{r} \right) \left( \langle \textbf{S}_2  \rangle_\xi \cdot \textbf{r} \right) + \nonumber \\ 
&+& 2 \left( \langle \textbf{S}_1  \rangle_\xi \cdot \textbf{r} \right) \left( \langle \textbf{S}_2  \rangle_\zeta \cdot \textbf{r} \right) \, ,\end{eqnarray} 
which also vanishes when $ \langle \textbf{S} \rangle_\zeta = 0$.

 In the next Section, we shall develop a prescription to extract a $4D-$potential from a $5D-$result. As we will see, this prescription enable us to bring some pseudo-spin contributions to $4D$.


\section{Restriction to 4D}
\label{S_restriction}
\indent


 To go over into a four-dimensional scenario, one may consider many different procedures, which are all based on at least  one ansatz. For example, a usual case assumes compactified extra-dimensions which lead to the Kaluza-Klein expansion modes. Also, we could impose a trivial reduction \cite{1975_Scherk_Schwarz}, where only field configurations that do not depend on the extra-dimensions are considered and only the so-called zero-modes are accounted for. Another possibility is to carry out the dimensional reduction by spontaneous compactification \cite{1976_Cremmer_Scherk}. In this case, we look for the solutions of the equations of motion which factorize into a four-dimensional space-time and an internal space.  On the other hand, one could also consider warped geometries - the so-called brane-worlds scenarios - in which the extra-dimensions are non-compact \cite{1999_Randall_Sundrum}. We also highlight a dimensional reduction prescription that does not assume compactified extra-dimensions neither take dynamical solutions.  That is the case of the Legendre reduction \cite{1980_Sohnius_Stelle_West}, normally adopted for  the construction of off-shell supersymmetric models.


After we have presented and discussed our results for the inter-particle potentials directly in $5D$, our purpose in the present Section is to carry out the dimensional reduction to $4D$. Nevertheless, instead of reducing the action and re-deriving the potentials in four dimensions from the reduced action, we pursue the attainment of the four-dimensional potentials by directly reducing the expressions calculated in five dimensions in the previous Section. So, we adopt the viewpoint already alluded to in the Introduction of our paper: the quantum-mechanical calculation – in our case is a semi-classical derivation of the potential and it is performed in the higher dimension to be, after that, reduced to the lower dimension. We proceed along the lines of the works quoted in refs. \cite{2006_Alvarez_Faedo}-\cite{1983_Alppequist_Chodos_Myers}, aiming at a four-dimensional result that already brings the semi-classical imprints from the five-dimensional physics. And what we actually conclude is that this procedure differs from the scheme of firstly reducing the action to then derive the potential from the reduced action. We converge to the claims of the papers by \'Alvarez and Faedo \cite{2006_Alvarez_Faedo}, who state that the two paths (reduction of the action followed by the inclusion of quantum effects or, alternatively, quantum effects worked out in the higher dimension to then reduce the quantum-corrected quantities to lower dimensions) may not be equivalent. We shall refer to the procedure we follow here as  reduction by dimensional restriction.

By inspecting the canonical mass dimension of the fields and coupling constants in $5D$ and $4D$, we have 
$ [\psi_{(5)}] = 2 \, , \, [A_{(5) \hat{\mu} } ] = 3/2 \, , \, [ g_{(5)} ] = -1/2 $ and 
$ [\psi_{(4)}] = 3/2 \, , \, [A_{(4) \mu } ] = 1 \, , \, [ g_{(4)} \equiv e ] = 0 $, respectively. 
If $L  $ denotes a length in the extra-dimension, then the factor $\sqrt{L}$  restores the correct mass dimension in $4D$,  such that $ g_{(4)} = g_{(5)} / \sqrt{L} $, $ \psi_{(4)} = \psi_{(5)} \, \sqrt{L} $ and
$ A_{(4) \hat{\mu} } = A_{(5) \hat{\mu} }\, \sqrt{L} $. For our purposes, we only need the relation between the coupling constants, which is independent of the dimensional reduction scheme.

Now, we propose the following procedure. First, we define the  average in the extra dimension of the potential in $5D$,
\begin{equation}  \langle V_{5D} \rangle_L = \frac{1}{L} \, \int_{-L/2}^{L/2} d \textbf{x}_4 \, V_{5D} \, , \label{def_res_0} \end{equation}
and then we extend this to a non-compact case, by taking the limit $L \rightarrow \infty$, 
\begin{equation}  V_{\textrm{res}} = \lim_{L \rightarrow \infty} \langle V_{5D} \rangle_L 
\, .\label{def_res} \end{equation} 

We refer to the potential $V_{\textrm{res}}$, defined in the previous equation, as a restricted potential or a restriction of the $5D-$potential to $4D$ space-time. 

In principle, one could think that this prescription is only
a statistical procedure, since in eq. \eqref{def_res_0} we have an average in the box $-L/2 < \textbf{x}_4 < L/2$ with equal probability $1/L$ and after we take the limit of a non-compact box. However, we shall also give an physical meaning to this.  If we substitute  eq. \eqref{def_pot_5D} in eq. \eqref{def_res_0}, we have that eq. \eqref{def_res} can be recast as
\begin{eqnarray}
V_{\textrm{res}}  = \lim_{L \rightarrow \infty} \, \frac{1}{L} \, \int_{-L/2}^{L/2} d \textbf{x}_4 \left\{ 
- \int \frac{d \textbf{q}_4}{2 \pi} \int \frac{d^3\textbf{q}}{(2\pi)^3} \, e^{i \textbf{q}_4 \textbf{x}_4} 
e^{i \textbf{q}\cdot \textbf{r}} \, \M_{NR}^{5D}\left[g_{(5)} \right] \, \right\}
\label{def_res_star} \end{eqnarray}

 In what follows, we shall use the relation $g_{(5)} = \sqrt{L} \, g_{(4)}$ between the fermionic coupling constants in $5D$ and $4D$. Before doing that, let us recall that the non-relativistic  tree-level amplitude  is proportional to the square of $g_{(5)}$ (see, for example, eq. \eqref{M_NR_5D}), so that the factor $1/L$ cancels against a factor coming from the coupling constants in $ \M_{NR}^{5D}\left[g_{(5)} \right] = L \, \M_{NR}^{5D}\left[g_{(4)} \right] $.  Next, we take the limit $L \rightarrow \infty$ and interchanging the integrals over $\textbf{x}_4$ and spatial momentum; this yields:
\begin{eqnarray}
V_{\textrm{res}}  &=& - \int \frac{d^3\textbf{q}}{(2\pi)^3} \, \int \frac{d \textbf{q}_4}{2 \pi} \left(  \int d \textbf{x}_4  e^{i \textbf{q}_4 \textbf{x}_4} \right) \, e^{i \textbf{q}\cdot \textbf{r}} \, \M_{NR}^{5D}\left[g_{(4)} \right] = \nonumber \\
&=&  - \int \frac{d^3\textbf{q}}{(2\pi)^3} \, \int d \textbf{q}_4 \, \delta(\textbf{q}_4)\, \, e^{i \textbf{q}\cdot \textbf{r}} \, \M_{NR}^{5D}\left[g_{(4)} \right] \, ,
\label{def_res_star_2} \end{eqnarray}
or, equivalently, 
\begin{eqnarray}
V_{\textrm{res}}  = - \int \frac{d^3\textbf{q}}{(2\pi)^3} \,   e^{i \textbf{q}\cdot \textbf{r}} \, \left[ \M^{5D}_{NR} \bigg|_{\textbf{q}_4 = 0} \right] \, 
\label{def_res_2} \end{eqnarray}
 where it is implicit  $g_{(5)} \rightarrow g_{(4)} $ in the above amplitude. Eq. \eqref{def_res_star_2} highlights that our reduction prescription naturally leads to $\textbf{q}_4 = 0$, in view of the Dirac delta function which comes out upon integration over $\textbf{x}_4$. 

Once in eq. \eqref{def_res_2} we take $\textbf{q}_4 = 0$, we could read the prescription as a restriction of the interaction to a subspace of the $5D$ space-time, without loss of the properties of the particles in $5D$, namely,  $\Delta$, (pseudo-)spin $\langle \textbf{S}^{\pm}  \rangle $ and momentum $\textbf{p} $, $\textbf{p}_4 $. For this reason, we shall avoid the expression  dimensional reduction. This prescription is just a restriction to the scattering amplitude, in which the transfer momentum of the extra dimension, $\textbf{q}_4$, could be considered negligible comparing to $\textbf{q}$ in the process. Remember that we are considering an elastic scattering, so we also have $q^0=0$. Here, we highlight that we assumed $\textbf{r} \neq 0$, so the restricted potential will not contemplate contact terms, i.e., ones with $\delta^{3}(\textbf{r}) $. To go over into  eq. \eqref{def_res_2}, we interchanged integrals and assumed non-singular functions.


 In our procedure, we draw attention to the fact that, by taking $\textbf{q}_4 = 0$, we are not setting the fifth component of the individual momenta to zero; in other words, we do not disregard the dependence of the fields on the extra space coordinate, $\textbf{x}_4$. What is zero here is the fifth component of the momentum transfer: the interaction of the matter currents with the intermediate boson does not transfer momentum along the fourth spatial component $(\textbf{q}_4)$ of the momentum transfer.  On the other hand, on the basis of our assumption given by
eq. \eqref{def_res} to carry out the dimensional reduction, the average taken over the extra-dimension goes from minus to plus infinity; this means that we are neither dropping out the $\textbf{x}_4-$dependence nor assuming $\textbf{x}_4$
to be compact. We are rather adopting the viewpoint of a non-compact extra-dimension, along
the lines of the idea proposed by Randall and Sundrum in the works \cite{1999_Randall_Sundrum}. This is the true reason why we do not consider the influence of the Kaluza-Klein tower of massive states.  Perhaps, we should also stress that, by considering the plane wave solutions given in eqs. \eqref{psi_p} and \eqref{chi_p}, we are already anticipating that non-compact dimensions will be present in our approach, which also confirms that Kaluza-Klein massive states are not considered here. 

 Though it is not our case in the present work, we would like to point out that Kaluza-Klein massive states, which appear as a consequence of the compactness of the extra-dimension, are of a very high mass and, at the energy compatible with the calculation of (low-energy) inter-particle potentials, they may be fairly-well disregarded. Actually,  they decouple. The momenta transfer in this sort of considerations are 
very low to excite the massive Kaluza-Klein states associated to the compact extra-dimensions. If these states were present, they would contribute as virtual particles running inside the momentum-space loop integrals that appear in the radiative corrections.

 Another point that we wish to stress is that, once the extra-coordinate $\textbf{x}_4$ is non-compact 
(which becomes explicit when we take the limit in eq. \eqref{def_res}) is that, through the Uncertainty
Principle, by fixing $\textbf{q}_4 = 0$, which is our basic assumption, we completely loose the localization
on $\textbf{x}_4$; this supports our prescription of taking the limit $L \rightarrow \infty$ of the  eq. \eqref{def_res_0}. All possible values of
$\textbf{x}_4$ are allowed (complete uncertainty on $\textbf{x}_4$) once $\textbf{q}_4 = 0$; this supports our prescription of taking the average on $\textbf{x}_4$.


 The procedure of taking integral in the extra dimension is not exclusive to this work. In ref. \cite{Arshansky_et_al}, the authors applied this integration in the fields and currents $-$ they called it a concatenation $-$ and this prescription was used in the Off-Shell Electrodynamics (see, for example, ref. \cite{off-shell_QED}). In our case, a different point of view is adopted. First, we carry out the inter-particle potential in $5D$, taking into account all the contribution of the fields and currents in $5D$, and then we integrate the potential.

Now let us apply the prescription to the $V^{\textrm{Maxwell}}_{5D} $, i.e., we  use eq. \eqref{def_res} and integrate eq. \eqref{EM_5D}, or, equivalently, we take the $3D$ Fourier integral of the amplitude, eqs. \eqref{M_NR_5D} and \eqref{JJ_5D},  with $\textbf{q}_4 =0$, which leads to the following result (for $\textbf{r} \neq 0$):
\begin{eqnarray} 
V^{\textrm{Maxwell}}_{\textrm{res}} & = & \frac{ e_1 e_2}{4 \pi r} \left\{ \Delta_1\Delta_2  \left ( 1 +  \frac{\textbf{p}^{2} + \textbf{p}_4^{2}}{m_1 m_2}  \right)   
+ \right. \nonumber\\ 
&-&  \frac{\textbf{L}}{ r^2} \,   \cdot \left[  
\Delta_1 \langle \textbf{S}^{+}_2 \rangle \left(\frac{1}{2 m_2^2} + \frac{1}{m_1 m_2} \right) + 1 \leftrightarrow 2 \right]   + \nonumber\\
&+&   \frac{\textbf{Q}_{ij}}{ r^2} \,\frac{1}{m_1 m_2} \,
\left[ \langle \textbf{S}^{+}_{1,i} \rangle \, \langle \textbf{S}^{+}_{2,j} \rangle - 
\langle \textbf{S}^{-}_{1,i} \rangle \, \langle \textbf{S}^{-}_{2,j}  \rangle \right] + \nonumber\\
& + &  \left. \frac{  \textbf{p}_4 \, \textbf{r} }{ r^2} \, \cdot \left[ \Delta_1 \langle \textbf{S}^{-}_2 \rangle \left(\frac{1}{2 m_2^2} + \frac{1}{m_1 m_2} \right) + 1 \leftrightarrow 2 \right]   \right\} 
\label{V_red} \end{eqnarray}

By comparing $V^{\textrm{Maxwell}}_{\textrm{res}}$ with  $V^{\textrm{Maxwell}}$, calculated directly in $4D$, eq. \eqref{V_4D_Maxwell} , we note some similarities after using the following dictionary: $\delta \leftrightarrow \Delta $ and $ \langle \textbf{S}  \rangle \leftrightarrow \langle \textbf{S}^{+}  \rangle $. We do not obtain modifications  for the  Coulomb term and, due our approximations, we do not also have interactions that couple $\textbf{S}^{+}$ with $\textbf{S}^{-} $. The $\textbf{p}_4 $ contribution appears coupled to pseudo-spin in a way similar to spin-orbit coupling, $  \textbf{L} \cdot \langle \textbf{S}^+  \rangle $. We highlight a new contribution to the quadrupole term, namely,  a pseudo-spin interaction, proportional to $\textbf{Q}_{ij} \, \langle \textbf{S}^{-}_{1,i} \rangle \, \langle \textbf{S}^{-}_{2,j} \rangle /r^2$. This new contribution is related to the interaction intermediate by the extra component of the $A^{\hat{\mu}}$. Even if ${\bm q}_4 = 0$, we have some contributions from the current, eq. \eqref{J4_2x_red}, which exhibits the coupling $\textbf{q} \cdot \langle \textbf{S}^{-} \rangle$.

 For all dimensional reduction schemes, the main requirement should be that the dominant contribution to the potential at large distances in the reduced four space-time dimensions, namely, the monopole-monopole interaction, decays with $r^{-1}$ and respects the well-known Coulomb's law \cite{teste_Coulomb_1} \cite{teste_Coulomb_2}. In our prescription, however, we arrive at a Coulomb potential and obtain an extra contribution to the quadrupole term due the presence of the pseudo-spin. We highlight that this is an effect driven by our reduction prescription.
It is worthy to decompose the (pseudo-)spin contributions to the quadrupole interaction in terms of the expectation values $\langle \textbf{S} \rangle_{\xi}$ and $\langle \textbf{S} \rangle_{\zeta}$ (see definitions in eq. \eqref{exp_values}). By using eq. \eqref{Spin_2x}, we obtain 

\begin{equation}
\frac{\textbf{Q}_{ij}}{ 4 \pi r^3} \,\frac{ 
 \langle \textbf{S}^{+}_{1,i} \rangle \, \langle \textbf{S}^{+}_{2,j} \rangle - 
\langle \textbf{S}^{-}_{1,i} \rangle \, \langle \textbf{S}^{-}_{2,j}  \rangle  }{m_1 m_2} = 
2 \, \frac{\textbf{Q}_{ij}}{ 4 \pi r^3} \,\frac{ 
 \langle \textbf{S}_{1,i} \rangle_\xi \, \langle \textbf{S}_{2,j} \rangle_\zeta + 
\langle \textbf{S}_{1,i} \rangle_\zeta \, \langle \textbf{S}_{2,j}  \rangle_\xi  }{m_1 m_2} \, .
\label{recast_quad} \end{equation}

 Hence, the quadrupole interaction appears only as  couplings between $\langle \textbf{S} \rangle_{\xi}$ and $\langle \textbf{S} \rangle_{\zeta}$. For this reason, in the parity-breaking case $(\zeta \rightarrow 0)$, this interaction vanishes.

\section{Proca Electrodynamics in 5D}
\label{S_Proca}
\indent

In the previous Section, we have discussed the inter-particle potential for the Maxwell Electrodynamics in $5D$. We wish now to generalize our results for a massive boson intermediate, described by the Proca Lagrangian,
\begin{equation} 
\L_{\textrm{Proca}} = - \frac{1}{4} \, F_{\hat{\mu} \hat{\nu}}^2 + \frac{1}{2} \, m^2 \, 
A_{\hat{\mu}}^2 \, . \label{L_Proca} 
\end{equation}

The propagator is given by
\begin{equation} 
\langle A_{\hat{\mu}} A_{\hat{\nu}} \rangle = - \frac{i}{q^2 - m^2} \left( \eta_{\hat{\mu} 
\hat{\nu}} - \frac{ q_{\hat{\mu}} q_{\hat{\nu}} }{m^2} \right) \, . \label{prop_Proca} 
\end{equation}

In a similar way done in Section \ref{S_Methodology}, we arrive at  $ \M = i \,  g_1 g_2 \, J_{(1)}^{\hat{\mu}} \, \langle  A_{\hat{\mu}} A_{\hat{\nu}} \rangle \, J_{(2)}^{\hat{\nu}}  $.  After using  the relation 
between $\M$  and $\M_{NR}$, eq. \eqref{def_Amp_NR}, we have 

\begin{equation}
\M_{NR}^{5D} = - \frac{g_1 g_2}{ \textbf{q}^{2} + \textbf{q}_4^{2} + m^2} \, \frac{J_{(1)}^{\hat{\mu}} J_{(2) \,\hat{\mu}}}{ (2E_1) (2E_2) } \, ,\label{M_Proca_5D} \end{equation}
where, in the last step, we have used the current conservation and $q^0 = 0$.

Note that the current contraction, presented in the last expression was carried out in eq. \eqref{JJ_5D}. Thus, considering the prescription in eq. \eqref{def_pot_5D} and using the Fourier integrals, eqs. \eqref{int_5D}-\eqref{int_5D_qq} in the Appendix, we obtain
%
\begin{eqnarray}
V^{\textrm{Proca}}_{5D} & = & g_1 g_2 \, \Biggl\{ \frac{m K_1}{4 \pi^2 R}  \, \Delta_1\Delta_2  \left[ 1 + \frac{\textbf{p}^{2}}{m_1 m_2} + \frac{\textbf{p}_4^{2}}{m_1 m_2} + \frac{m^2}{8} \left( \frac{1}{m_1^2} + \frac{1}{m_2^2}\right) \right]  + 
\nonumber\\
&-&  \frac{\Delta_1\Delta_2}{8}   \left( \frac{1}{m_1^2} + \frac{1}{m_2^2} \right)  \delta^{4}(\textbf{R}) 
- \frac{ \, \langle \textbf{S}^{+}_1 \rangle \cdot \langle \textbf{S}^{+}_2 \rangle + 
\langle \textbf{S}^{-}_1 \rangle \cdot \langle \textbf{S}^{-}_2  \rangle \, }{2 m_1 m_2} \delta^{4}(\textbf{R}) + \nonumber\\
& + &  \frac{m^3 K_1}{4 \pi^2 R} \, \frac{\langle \textbf{S}^{-}_1 \rangle \cdot \langle \textbf{S}^{-}_2 \rangle}{m_1 m_2}   
-  \frac{1}{4\pi^2 R^2}\left[ \frac{4m K_1}{R} + 2 m^2 K_0 
\right. + \nonumber\\
&-& \left. \frac{r^2}{R} \left( \frac{8m K_1}{R^2} + \frac{4m^2 K_0}{R} + m^3 K_1 \right) \right] 
 \frac{  \langle \textbf{S}^{+}_1 \rangle \cdot \langle \textbf{S}^{+}_2 \rangle - \langle \textbf{S}^{-}_1 \rangle \cdot \langle \textbf{S}^{-}_2  \rangle }{  m_1 m_2} + \nonumber\\
& - & \frac{1}{4 \pi^2 R^2} \left( \frac{2m K_1}{R} + m^2 K_0 \right) \left[ \left( \textbf{r} \times \textbf{p} \right) \cdot \left(  
\Delta_1 \langle \textbf{S}^{+}_2 \rangle \left(\frac{1}{2 m_2^2} + \frac{1}{m_1 m_2} \right)
\right)  \right. + \nonumber\\ 
& + & \left. \left( \textbf{x}_4 \, \textbf{p} - \textbf{p}_4 \, \textbf{r} \right) \cdot \left( \Delta_1 \langle \textbf{S}^{-}_2 \rangle \left(\frac{1}{2 m_2^2} + \frac{1}{m_1 m_2} \right) \right) + 1 \leftrightarrow 2 \right] + \nonumber\\
& - &  \frac{1}{4 \pi^2 R^3} \left( \frac{8m K_1}{R^2} + \frac{4 m^2 K_0}{R} + m^3 K_1 \right) \frac{1}{m_1 m_2} \biggl[ \textbf{x}_4 \, \textbf{r} \cdot \left[ \, \left( \langle \textbf{S}^{+}_1 \rangle \times \langle \textbf{S}^{-}_2 \rangle \right) + \right.  \nonumber\\
&+&  \left. \left( \langle \textbf{S}^{+}_2 \rangle \times \langle \textbf{S}^{-}_1 \rangle \right) \right] 
+  \left( \langle \textbf{S}^{+}_1  \rangle \cdot \textbf{r} \right) \left( \langle \textbf{S}^{+}_2  \rangle \cdot \textbf{r} \right)  
-    \left( \langle \textbf{S}^{-}_1  \rangle \cdot \textbf{r} \right) \left( \langle \textbf{S}^{-}_2  \rangle \cdot \textbf{r} \right) \biggr] \Biggr\} \, .
\label{Proca_5D} \end{eqnarray} 

This potential exhibits all the velocity- and (pseudo)-spin-dependence interactions of the Maxwell case, eq.\eqref{EM_5D}, which is recovered in the massless limit. In the Proca potential, the power-law decay depends on the  Modified Bessel function, $K_\nu(z)$, and  the  range of $z=mR$. In the sequel, we shall study its asymptotic behaviors as $R$ goes to infinity and zero, respectively, and then present its form upon the reduction by dimensional restriction.
 


According to refs. \cite{Abramowitz_Stegun} \cite{K_e_z_inf} , the behavior of the  $K_\nu (z)$, when $z \rightarrow \infty $ is given by 
\begin{equation}
K_\nu (z) \sim \sqrt{\frac{\pi}{2z}} \, e^{-z} \, \left[ 1 + O\left( \frac{1}{z} \right) \right] \, , 
\label{K_inf} \end{equation}
which holds  for $ | \arg z | < 3\pi/2$. Since we are using real values $z= m R$, the previous condition is automatically satisfied.
Applying this result in eq. \eqref{Proca_5D}, we obtain the following asymptotic limit   
\begin{eqnarray}
V^{\textrm{Proca}}_{5D} \biggl|_{R \rightarrow \infty} & \sim & \frac{g_1 g_2}{4} \, \sqrt{\frac{m}{2 \pi^3}} \, e^{- mR} \left\{ 
\frac{ \Delta_1 \Delta_2}{R^{3/2}}  \left(1 + \frac{\textbf{p}^{2}}{m_1 m_2} + \frac{\textbf{p}_4^{2}}{m_1 m_2} \right) \right. + \nonumber\\ 
 & + & \frac{m^2}{8} \, \frac{ \Delta_1 \Delta_2}{R^{3/2}}  \left( \frac{1}{m_1^2} + \frac{1}{ m_2^2} \right)
 \, + \frac{ m^2}{m_1 m_2} \frac{1}{R^{3/2}} \, \langle \textbf{S}^{-}_1 \rangle \cdot \langle \textbf{S}^{-}_2 \rangle  + \nonumber\\  
& - &  m \, \frac{ \left( \textbf{r} \times \textbf{p} \right) }{R^{5/2}} \cdot \left[  
\Delta_1 \langle \textbf{S}^{+}_2 \rangle \left(\frac{1}{2 m_2^2} + \frac{1}{m_1 m_2} \right)
 + 1 \leftrightarrow 2 \right] + \nonumber\\ 
& - & m \, \frac{ \left( \textbf{x}_4 \, \textbf{p} - \textbf{p}_4 \, \textbf{r} \right) }{R^{5/2}} \, 
 \cdot \left[ \Delta_1 \langle \textbf{S}^{-}_2 \rangle \left(\frac{1}{2 m_2^2} + \frac{1}{m_1 m_2} \right)  + 1 \leftrightarrow 2 \right] + \nonumber\\ 
 & - & \frac{m^2}{m_1 m_2} \, \frac{\textbf{x}_4 \, \textbf{r}}{R^{7/2}} \cdot \left[ \left( \langle \textbf{S}^{+}_1 \rangle \times \langle \textbf{S}^{-}_2 \rangle \right) + \left( \langle \textbf{S}^{+}_2 \rangle \times \langle \textbf{S}^{-}_1 \rangle \right) \right]  + \nonumber\\ 
& + & \frac{m^2}{m_1 m_2} \, \frac{r^2}{R^{7/2}} \,  \left[ \langle \textbf{S}^{+}_1 \rangle \cdot \langle \textbf{S}^{+}_2 \rangle - 
\langle \textbf{S}^{-}_1 \rangle \cdot \langle \textbf{S}^{-}_2  \rangle \right]
 + \nonumber\\ 
 & - & \left. \frac{m^2}{m_1 m_2} \, \frac{1}{R^{7/2}} \, \left[  \left( \langle \textbf{S}^{+}_1  \rangle \cdot \textbf{r} \right) \left( \langle \textbf{S}^{+}_2  \rangle \cdot \textbf{r} \right)
- \left( \langle \textbf{S}^{-}_1  \rangle \cdot \textbf{r} \right) \left( \langle \textbf{S}^{-}_2  \rangle \cdot \textbf{r} \right) \right] \right\} \, .
\label{Proca_5D_R_inf} \end{eqnarray} 

Here, we notice a peculiar behavior. In addition to the common factor $e^{-mR}$, we also have  fractional power-law decay in all terms. The dominant monopole-monopole contribution decays with $R^{3/2}$.  The  velocity- and (pseudo-)spin-dependent terms are suppressed by the mass fermions and have higher power law decay.

Let us now consider the situation when $z \rightarrow 0$. 
The asymptotic limits are given by refs. \cite{K_0_e_z_0} and \cite{K_1_e_z_0}, respectively, %
\begin{equation}
K_0 (z) \sim - \log\left( \frac{z}{2} \right) \, \left[ 1 + O(z^2) \right] - \gamma \left[ 1 + O(z^2) \right] \, , \label{K_0_tend_0} \end{equation}
\begin{equation}
K_1 (z) \sim  \frac{1}{z} \left[ 1 + O(z^2) \right] + \frac{z}{2} \log \left( \frac{z}{2} \right) \, \left[ 1 + O(z^2) \right] \, ,
\label{K_1_tend_0} \end{equation}
where $\gamma = 0.57721...$ is the Euler-Mascheroni constant.

By taking into account these limits in eq. \eqref{Proca_5D}, one may obtain
\begin{eqnarray}
V^{\textrm{Proca}}_{5D} \biggl|_{R \rightarrow 0} & \sim & \frac{g_1 g_2}{4 \pi^2 R^2} \, 
\Biggl\{ \Delta_1\Delta_2   \,  \left[ 1 + \frac{\textbf{p}^{2}}{m_1 m_2} + \frac{\textbf{p}_4^{2}}{m_1 m_2} + \frac{m^2}{8} \left( \frac{1}{m_1^2} + \frac{1}{m_2^2}\right) \right]  + 
\nonumber\\
& + &  \frac{m^2 }{m_1 m_2} \, \langle \textbf{S}^{-}_1 \rangle \cdot \langle \textbf{S}^{-}_2 \rangle   
-  \left[ \frac{4}{R^2} - 2 m^2 \left( \log\left( \frac{mR}{2}\right) + \gamma \right)
\right. + \nonumber\\
&-& \left. \frac{r^2}{R^2} \left( \frac{8}{R^2} + m^2 \left( 1 - 4 \gamma -4 \log\left( \frac{mR}{2}\right) \right) \right) \right] 
 \frac{  \langle \textbf{S}^{+}_1 \rangle \cdot \langle \textbf{S}^{+}_2 \rangle - \langle \textbf{S}^{-}_1 \rangle \cdot \langle \textbf{S}^{-}_2  \rangle }{  m_1 m_2} + \nonumber\\
& - &  \left( \frac{2}{R^2} - m^2 \left( \log\left( \frac{mR}{2}\right) + \gamma \right) \right) \left[ \left( \textbf{r} \times \textbf{p} \right) \cdot \left(  
\Delta_1 \langle \textbf{S}^{+}_2 \rangle \left(\frac{1}{2 m_2^2} + \frac{1}{m_1 m_2} \right)
\right)  \right. + \nonumber\\ 
& + & \left. \left( \textbf{x}_4 \, \textbf{p} - \textbf{p}_4 \, \textbf{r} \right) \cdot \left( \Delta_1 \langle \textbf{S}^{-}_2 \rangle \left(\frac{1}{2 m_2^2} + \frac{1}{m_1 m_2} \right) \right) + 1 \leftrightarrow 2 \right] + \nonumber\\
& - &  \frac{1}{ R^2} \left( \frac{8}{R^2} + m^2 \left( 1 - 4 \gamma -4 \log\left( \frac{mR}{2}\right) \right)  \right) 
\frac{1}{m_1 m_2} \biggl[ \textbf{x}_4 \, \textbf{r} \cdot \left[ \, \left( \langle \textbf{S}^{+}_1 \rangle \times \langle \textbf{S}^{-}_2 \rangle \right) + \right.  \nonumber\\
&+&  \left. \left( \langle \textbf{S}^{+}_2 \rangle \times \langle \textbf{S}^{-}_1 \rangle \right) \right] 
+  \left( \langle \textbf{S}^{+}_1  \rangle \cdot \textbf{r} \right) \left( \langle \textbf{S}^{+}_2  \rangle \cdot \textbf{r} \right)  
-    \left( \langle \textbf{S}^{-}_1  \rangle \cdot \textbf{r} \right) \left( \langle \textbf{S}^{-}_2  \rangle \cdot \textbf{r} \right) \biggr] \Biggr\} \, .
\label{Proca_5D_z=0} \end{eqnarray} 
%

 Next, we present the restriction to $4D$ of the Proca Electrodynamics studied in $5D$. Following the prescription discussed in Section \ref{S_restriction}, i.e., we take the limit $\textbf{q}_4 \rightarrow 0$ in the non-relativistic amplitude, eq. \eqref{M_Proca_5D} with eq. \eqref{JJ_5D}, and work out the Fourier integrals in $3D$ for $r \neq 0$ (see eqs. \eqref{int_4D}-\eqref{int_4D_qq} in Appendix \ref{Apendice}), which leads to

\begin{eqnarray} 
V^{\textrm{Proca}}_{\textrm{res}} & = & e_1 e_2 \frac{ e^{-mr} }{4 \pi r}  \, \left\{ \Delta_1\Delta_2 \left[  \left ( 1 +  \frac{\textbf{p}^{2} + \textbf{p}_4^{2}}{m_1 m_2}  \right) 
\right. \right] + \frac{m^2}{m_1 m_2} \, \langle \textbf{S}^{+}_1 \rangle \cdot \langle \textbf{S}^{+}_2 \rangle  \,  + \nonumber \\
&-&  \textbf{L}\, \cdot \left[  
\Delta_1 \langle \textbf{S}^{+}_2 \rangle \left(\frac{1}{2 m_2^2} + \frac{1}{m_1 m_2} \right) + 1 \leftrightarrow 2 \right]  \frac{\left( 1 + m r \right)}{r^2} + \nonumber\\
&+&  \frac{1}{m_1 m_2}  \,  \frac{\textbf{Q}_{ij}^{(m)}}{r^2} \,
\left[ \langle \textbf{S}^{+}_{1,i} \rangle \, \langle \textbf{S}^{+}_{2,j} \rangle - 
\langle \textbf{S}^{-}_{1,i} \rangle \, \langle \textbf{S}^{-}_{2,j}  \rangle \right] \,   + \nonumber\\
& + &  \left.   \textbf{p}_4 \, \textbf{r}  \, \cdot \left[ \Delta_1 \langle \textbf{S}^{-}_2 \rangle \left(\frac{1}{2 m_2^2} + \frac{1}{m_1 m_2} \right) + 1 \leftrightarrow 2 \right] \frac{\left( 1 + m r \right)}{r^2}   \right\} \, ,
\label{V_Proca_red} \end{eqnarray}

where we  defined
\begin{equation}
\textbf{Q}_{ij}^{(m)} = (1+ mr) \, \delta_{ij} - (3 + 3mr + m^2 r^2) \,  \frac{ \textbf{x}_i \textbf{x}_j }{r^2} \, .
\label{def_quad_m} \end{equation}

It is worth to compare this potential with the Maxwell case, eq. \eqref{V_4D_Maxwell}. Again the pseudo-spin contributions appear in a coupling with $\textbf{p}_4 $ and in the quadrupole term.

 Finally, let us discuss an illustrative case in which we do not take $\textbf{q}_4 =0$, but we consider small contributions. For the sake of simplicity, we assume the monopole-monopole interaction,  described by the following amplitude

\begin{equation}
\M_{NR}^{5D} = - \frac{g_1 g_2}{ \textbf{q}^{2} + \textbf{q}_4^{2} + m^2} \, \, \Delta_1 \Delta_2\,  .\label{M_NR_5D_small} \end{equation}

By plugging this amplitude into eq. \eqref{def_pot_5D} and carrying out the integration $ \int d^3 \textbf{q}$, one  arrives at
\begin{equation} V =   g_1 g_2 \, \Delta_1 \Delta_2   \, \int \frac{d 
\textbf{q}_4 
}{\left( 2 \pi \right)} \, e^{i \textbf{q}_4  \textbf{x}_4}  \, \left[ \frac{1}{4 
\pi r} \, e^{- m r \sqrt{ 1 + \textbf{q}^{\, 2}_4/m^2 }}  \right] \, . \end{equation}

 Now, if one consider $\textbf{q}_4^2 << m^2$, it is possible to approximate $e^{-mr \sqrt{1 +  \textbf{q}^{\,2}_4 / m^2 }} \approx 
e^{-mr} \, e^{-r \textbf{q}_4^{\,2}/2m  }$ and the  Fourier integral above reduces to a Gaussian one, which, for large distances (with $ m r >> m^2 \textbf{x}_4^2 $), leads to the potential

\begin{eqnarray}  V & \approx &   g_1 g_2 \, \left[  \frac{\Delta_1 \Delta_2}{4 
\pi r} \, e^{-mr}  
\right] \sqrt{\frac{m}{2 \pi r}} \left( 1 - \frac{m^2 \textbf{x}_4^2}{2mr} + ... \right) \, .
\label{V_1} \end{eqnarray}

From this example, we conclude that, if we take even a small $\textbf{q}_4-$contribution, the dominant term decays with $e^{-mr}/r^{3/2}$ rather than showing the usual Yukawa-like profile in $4D$, given by $e^{-mr}/r$. The term with $\textbf{x}_4-$dependence falls off with $e^{-mr}/r^{5/2}$. In our prescription, after imposing the dimensional restriction, we arrive at the condition $\textbf{q}_4 = 0$ and, consequently, obtain the expected Yukawa (or Coulomb in the massless case) dominant interaction in $4D$. Moreover, we have shown that new contributions appear in the spin sector due the presence of pseudo-spin as an inheritance of $5D$ space-time.

\section{Concluding Comments}
\label{S_Concluding}
\indent

As it has been discussed over the past Sections, our main endeavour in this paper was to compute photon-mediated and Proca-mediated parity-preserving inter-particle potentials in $5D$  to get their four-dimensional description by adopting a particular scheme, which we refer to as reduction by dimensional restriction. The main feature of this procedure is the prescription that the mediating particle (in the cases we considered, massless and massive abelian vector bosons) does not transfer momentum in the extra spatial dimension $(\textbf{q}_4 = 0)$. Our claim is that the physics of the interaction process exchanges momentum only along the $\textbf{q}_1, \textbf{q}_2$ and $\textbf{q}_3$ directions in momentum-transfer space  ($\textbf{q}_0 = 0$ for an elastic scattering). With this assumption, we correctly get the right space dependence of the potentials. Had we considered  a non-trivial momentum transfer along $\textbf{q}_4$, the dependence of the four-dimensional potentials would not be the right ones; they would fall off faster with distance (in the case of the monopole contribution, as an illustration, it would be $r^{-2}$  rather than  $r^{-1}$). 

The behavior of the potentials with the particles’ spatial separation, velocities and spins is also worked out in details in the tree-level approximation. And, as a consequence of setting up the physics in $5D$, there emerges, in $4D$, an extra degree of freedom that we here name pseudo-spin; that is not the same as the pseudo-spin that appears in other contexts, as we have previously pointed out.  Actually, the appearance of the pseudo-spin in our prescription seems to be a new feature and  we wish to go deeper into this point.  The potential obtained in eq. \eqref{V_red} and the decomposition of the quadrupole interaction in terms of two spins in $4D$, eq. \eqref{recast_quad}, are our departure to better understand the role of pseudo-spin in four-dimensional physical processes.  This quadrupole-type contribution is a non-trivial consequence of the scheme we are referring to as dimensional restriction. We have, in particular, already initiated to pursue a study of the pseudo-spin in connection with the multipole structure of the fermionic current, with particular attention to a possible relationship between pseudo-spin and the electron and muon electric dipole moments in models where there occurs $CP-$violation. We intend to report on that elsewhere in a forthcoming work.

It is worthy to mention that we have in this paper considered a particular way to introduce the fermion mass without breaking parity in five dimensions. We have doubled the fermion representation and defined parity in a particular way, by imposing that the fermions of the doublet are exchanged into one another upon the action of parity transformation. In connection with the doubling of the spinors that represent the fermion in five dimensions, we may introduce a number of different symmetries, as we highlight in the next paragraph. As a new possibility that opens up, it would be interesting to understand how these symmetries may affect four-dimensional physics, specially in association with the electron's and muon's electric and magnetic dipole moments. This shall be object of our immediate interest. 


In a similar way as it was done in $3D$ \cite{QED_3}, one may introduce other global (or local) phase transformations for the doubled spinor field in $5D$. These transformations are defined by  using $8 \times 8$ matrices, namely,
\begin{equation}
\tau_3 = \left( \begin{array}{cc}
 0 & 1 \\ 
 1 & 0\end{array} \right) \,  \; ,  \, \; 
 \tau_4 = i \left( \begin{array}{cc}
 0 & 1 \\ 
 - 1 & 0\end{array} \right) \, \; , \, \;
 \tau_5 = \left( \begin{array}{cc}
 1 & 0 \\ 
 0 & - 1\end{array} \right) \, .
\end{equation}

It is possible to show that the  massless term $\overline{\Psi} \, i \, \Gamma^{\hat{\mu}} \, \partial_{\hat{\mu}} \Psi$ is invariant under these transformations. However, the mass term breaks the $\tau_3$ and $\tau_4$ symmetries. Only the $\tau_5-$case is consistent with a mass term. Furthermore, if we consider a local $\tau_5-$symmetry, we obtain the  current $J_5^{\hat{\mu}} \equiv  \overline{\Psi} \,  \Gamma^{\hat{\mu}} \tau_5 \Psi$ and an Abelian gauge field, $B^{\hat{\mu}}$, both pseudo-vectors in $5D$. In this case, beyond the Maxwell-like term in the Lagrangian, we could also introduce a Chern-Simons term in $5D$, without breaking the parity symmetry. This particular case could be more explored in connection with topological superconductors \cite{Top_insulator}, where a Chern-Simons term  plays an important role.

Finally, we point out that applying the dimensional restriction prescription to go from $(1+3)-$ to a $(1+2)-$dimensional space-time may be of interest in the inspection of low-dimensional systems in Condensed Matter Physics, such as graphene and charge/spin Hall effect.


\begin{acknowledgments}

We would like to thank Tobias Micklitz for useful and critical comments and rich exchange of ideas. This work was supported by  the {\it Foundation for the Support to Research of the Rio de Janeiro State} (FAPERJ) and {\it National Council for Scientific
and Technological Development} (CNPq/MCTIC) through the PCI-DB funds.

\end{acknowledgments}


\appendix
\section{Fourier Integrals}
\label{Apendice}

Below, we present some useful Fourier Integrals in $3D$ and $4D$. Let us initiate with the well-known $3D$ massive case: 
\begin{equation}
\int  \frac{d^3 \textbf{q}}{(2 \pi)^3}  \, \frac{e^{i \textbf{q} \cdot \textbf{r}} }{\textbf{q}^{2}  + m^2} = \frac{e^{-mr}}{4 \pi r} \, , 
\label{int_4D} \end{equation}

\begin{equation}
\int  \frac{d^3 \textbf{q}}{(2 \pi)^3}   \, \frac{ e^{i \textbf{q} \cdot \textbf{r}}}{\textbf{q}^{2} + m^2} \, \textbf{q}_i \, = 
\frac{i \, \textbf{x}_i}{4 \pi r^3} \, \left( 1 + mr \right) \, e^{-mr} \, ,
\label{int_4D_q} \end{equation}
 
\begin{equation}
\int  \frac{d^3 \textbf{q}}{(2 \pi)^3}   \, \frac{e^{i \textbf{q} \cdot \textbf{r}}}{\textbf{q}^{2} + m^2} \, \textbf{q}_i \textbf{q}_j \, = \frac{\delta_{ij}}{3} \delta^3(\textbf{r}) + \frac{e^{-mr}}{4 \pi r^3} \, \left[ \left( 1 + mr \right) \delta_{ij} - \left( 3 + 3mr + m^2 r^2 \right) \frac{\textbf{x}_i \, \textbf{x}_j}{r^2} \right]
\, , \label{int_4D_qq} \end{equation}
where $ r = \sqrt{ \textbf{r}^2 }$ and $i,j=1,2,3$. From these equations one can directly obtain the massless limit.

In order to avoid confusion, we shall use $\textbf{R}$ to denote the $4D$ (euclidean) vector and $\textbf{x}_I$ for its components, with capital letter $I=1,2,3,4$.  The $4D$ massive Fourier integrals are given by  
\begin{equation}
\int  \frac{d^4 \textbf{q}}{(2 \pi)^4} \, \frac{e^{i \textbf{q} \cdot \textbf{R}}}{\textbf{q}^{2} + m^2} = \frac{m}{4 \pi^2 R} \, K_1(mR) \, ,
\label{int_5D} \end{equation}
 
\begin{equation}
\int  \frac{d^4 \textbf{q}}{(2 \pi)^4} \, \frac{  e^{i \textbf{q} \cdot \textbf{R}}}{\textbf{q}^{2} + m^2} \, \textbf{q}_I \, = 
\frac{i \, \textbf{x}_I}{4 \pi^2 R^2} \, \left[ \frac{2m K_1(mR)}{R} + m^2 K_0(mR) \right] \, , 
\label{int_5D_q} \end{equation}

\begin{eqnarray}
\int  \frac{d^4 \textbf{q}}{(2 \pi)^4} \, \frac{e^{i \textbf{q} \cdot \textbf{R}}}{\textbf{q}^{2} + m^2} \, \textbf{q}_I \, \textbf{q}_J  &=& 
\frac{1}{4} \, \delta_{IJ} \, \delta^4(\textbf{R}) + \frac{\delta_{IJ}}{4 \pi^2 R^2} \, \left[ \frac{2m K_1(mR)}{R} + m^2 K_0(mR) \right] + \nonumber\\
&-& \frac{\textbf{x}_I \textbf{x}_J}{4 \pi^2 R^3} \, 
\left[ \frac{8m K_1(mR)}{R^2} + \frac{4m^2 K_0(mR)}{R} + m^3 K_1(mR) \right] \, ,
\label{int_5D_qq} \end{eqnarray}
where $R = \sqrt{\textbf{R}^2} $ and $K_\nu(z)$ is the modified Bessel function of the second kind with order $\nu$.
  

By using the asymptotic limits, eqs. \eqref{K_0_tend_0} and \eqref{K_1_tend_0}, it is possible to work out the massless limits, which take the form:
\begin{equation}
\int  \frac{d^4 \textbf{q}}{(2 \pi)^4} \, \frac{e^{i \textbf{q} \cdot \textbf{R}}}{\textbf{q}^{2} } = \frac{1}{4 \pi^2 R^2} \, ,
\label{int_massless}
\end{equation}
\begin{equation}
\int  \frac{d^4 \textbf{q}}{(2 \pi)^4} \, \frac{e^{i \textbf{q} \cdot \textbf{R}}}{\textbf{q}^{2} } \, \textbf{q}_I = \frac{i}{2 \pi^2 }\frac{\textbf{x}_I}{R^4} 
\label{int_q_massless} \, ,
\end{equation}
\begin{equation}
\int  \frac{d^4 \textbf{q}}{(2 \pi)^4} \, \frac{e^{i \textbf{q} \cdot \textbf{R}}}{\textbf{q}^{2} } \, \textbf{q}_I  \, \textbf{q}_J = 
\frac{1}{4} \, \delta_{IJ} \, \delta^4(\textbf{R}) +
\frac{1}{2 \pi^2 R^4} \left[ \delta_{IJ} - 4 \frac{\textbf{x}_I \textbf{x}_J}{R^2} \right] \, .
\label{int_qq_massless}
\end{equation}

\end{document}